\documentclass[a4paper,12pt]{article}
\pdfoutput=1

\usepackage{graphicx}
\usepackage{mathptmx}      
\usepackage{latexsym}

\usepackage{amsmath}
\usepackage{amsfonts}
\usepackage{amssymb}
\usepackage{amsthm}

\usepackage[amssymb,Gray,derived,derivedinbase]{SIunits}
\usepackage[np,nosepfour]{numprint} 

\usepackage[T1]{fontenc}
\usepackage[utf8]{inputenc}
\usepackage{lmodern}
\usepackage{microtype}

\usepackage{indentfirst}
\usepackage{multicol}
\usepackage[shortlabels]{enumitem}
\usepackage{xfrac}

\usepackage[x11names,svgnames,table]{xcolor} 
\usepackage{ctable}

\usepackage{graphicx}
\usepackage[config, labelfont={normalfont,bf}, font={small}]{caption}
\usepackage[config, labelfont={normalfont,it}, font={small}]{subfig}

\usepackage[round,merge]{natbib}

\newcommand{\betti}[1]{\ensuremath{\mathcal{B}_{#1}}}

\newcommand{\med}{\ensuremath{_*}}

\newcommand{\e}[1]{\ensuremath{\allowbreak\times\allowbreak 10^{#1}}}

\newcommand{\E}[1]{\ensuremath{\mathrm{\,\textsc{e}}{#1}}}

\newcommand{\Rr}{\ensuremath{R(r)}}

\newcommand{\dtpYL}{PSD$_{YL}$}

\newcommand{\apendice}[1]{Ap. #1}
\newcommand{\bibCunhaMSA}{Cunha2013Physicae}

\newcommand{\plt}{}

\newcommand{\corE}{black}

\usepackage[%
        pdftex,
        colorlinks=true,
        breaklinks=true,
        bookmarks=true,
        bookmarksnumbered=true,
        bookmarksopen=true,%
        plainpages=false,
        breaklinks=true,
        anchorcolor=\corE,
        citebordercolor=\corE,
        citecolor=\corE,
        linkcolor=\corE,
        linkbordercolor=\corE,
        filebordercolor=\corE,
        menubordercolor=\corE,
        urlcolor=\corE,
        urlbordercolor=\corE,
        pdftitle={Conectividade em materiais porosos},
        pdfauthor={André Rafael Cunha},
        pdfsubject={Propriedades de transporte de materiais porosos},
        pdfkeywords={Meios Porosos; Conectividade; Permeabilidade Absoluta},
        pdflang=Portugues
]{hyperref}

\usepackage{breakurl}

\newcommand{\pastaFig}{.}
\newcommand{\pastaGraf}{.}

\newcommand{\sF}[1]{\textit{#1)}}

\newcommand{\figResultadoBM}{%
\begin{figure}[!h]
    \centering
    \fbox{\includegraphics[keepaspectratio=true,width=5cm]{\pastaFig/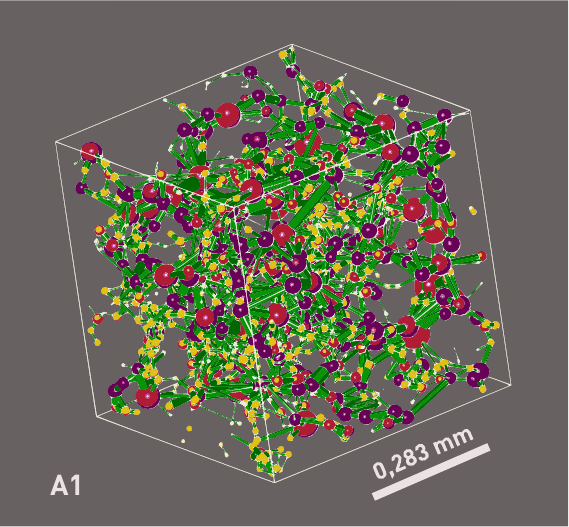}}
    \caption{Result of the application of the Maximum Ball Algorithm.}
    \label{Fig:ResultadoBM}
\end{figure}%
}

\newcommand{\grafDTPAren}{%
    \begin{figure}[!h]
        \centering
        \includegraphics[keepaspectratio=true,width=0.85\textwidth]{\pastaGraf/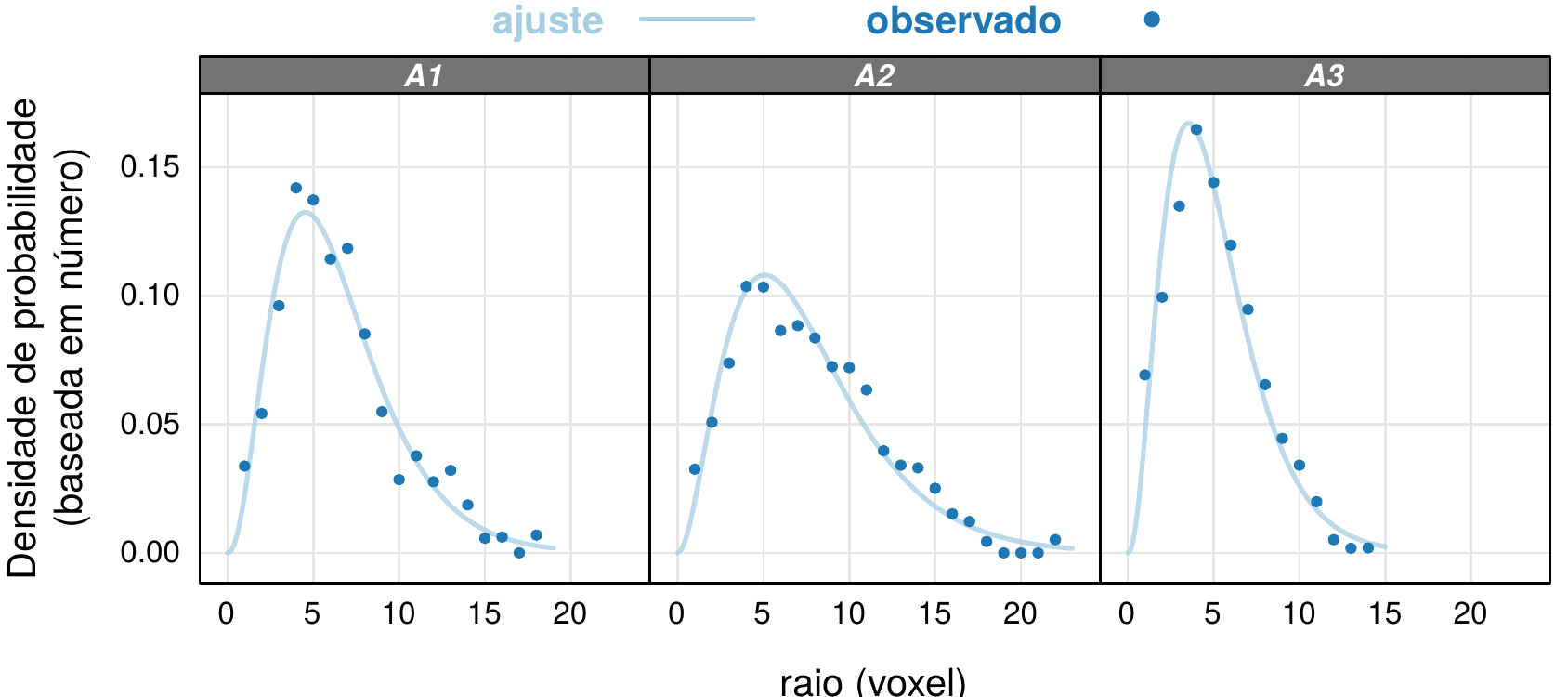}
        \caption{Spherical Pore Size Distribution (S-PSD) \Rr\ of sandstone samples fitted by the gamma distribution.}
        \label{Graf:DTPAren}
    \end{figure}%
}

\newcommand{\grafCorrelAren}{%
    \begin{figure}[!h]
        \centering
        \includegraphics[keepaspectratio=true,width=0.8\textwidth]{\pastaGraf/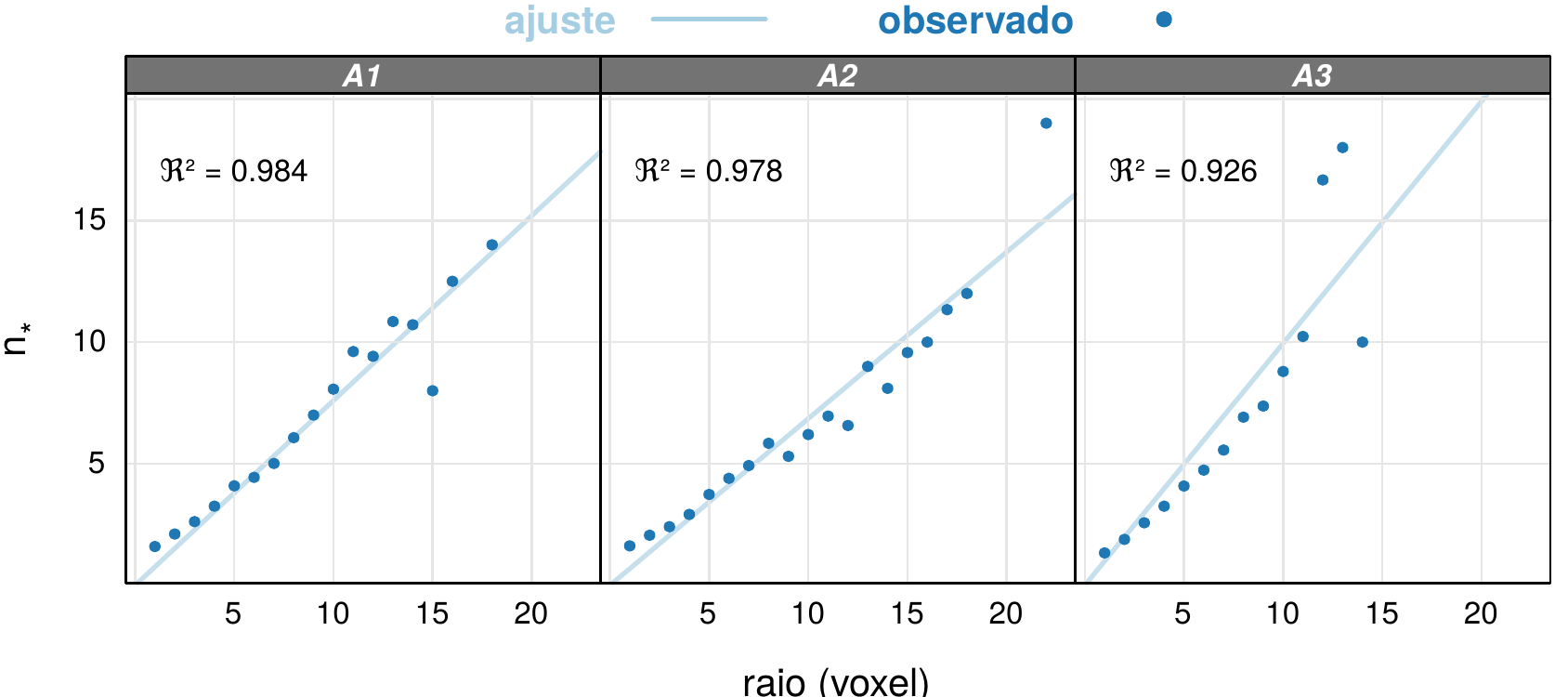}
        \caption{Correlation between mean coordination number $n\med$ of a pore and its radius $r$ for sandstone samples.}
        \label{Graf:CorrelAren}
    \end{figure}%
}

\newcommand{\grafCon}{%
    \begin{figure}[!h]
        \centering
        \includegraphics[keepaspectratio=true,width=0.45\textwidth]{\pastaGraf/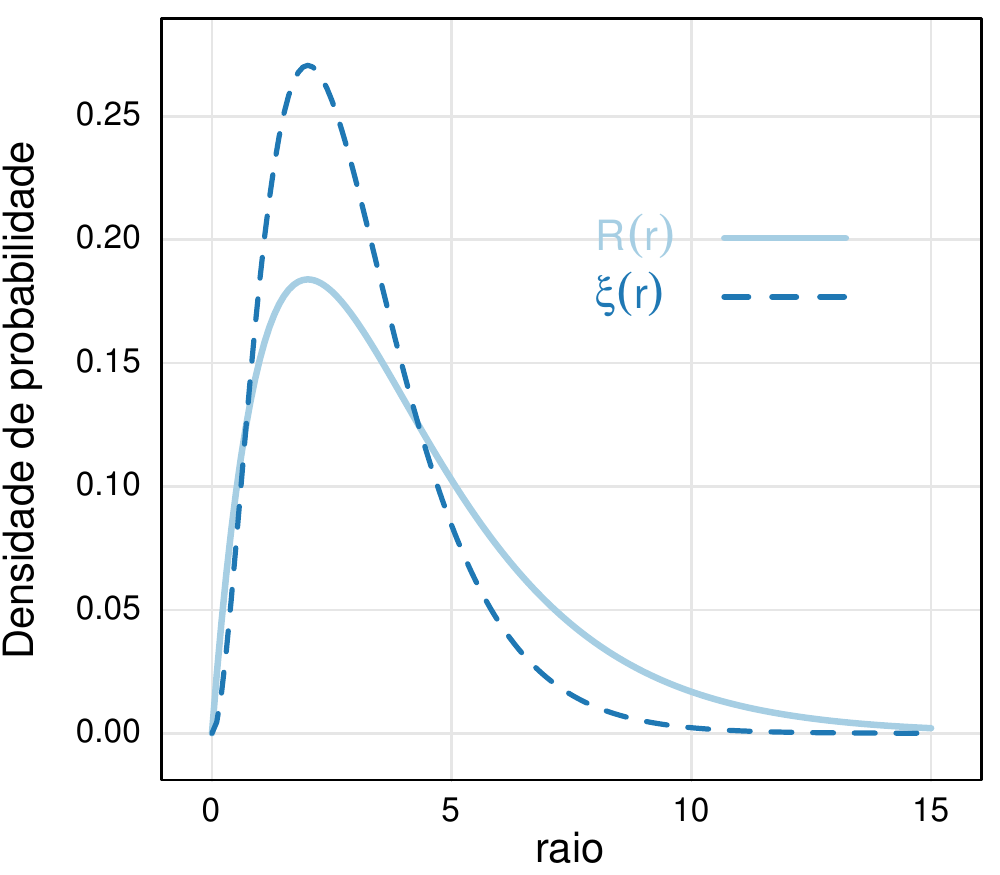}
        \caption{\Rr\ and $\xi(r)$ for $\alpha=2$ and $\beta=2$.}
        \label{Graf:Con}
    \end{figure}%
}

\newcommand{\grafSP}{%
    \begin{figure}[!h]
        \centering
        \includegraphics[keepaspectratio=true,width=0.8\textwidth]{\pastaGraf/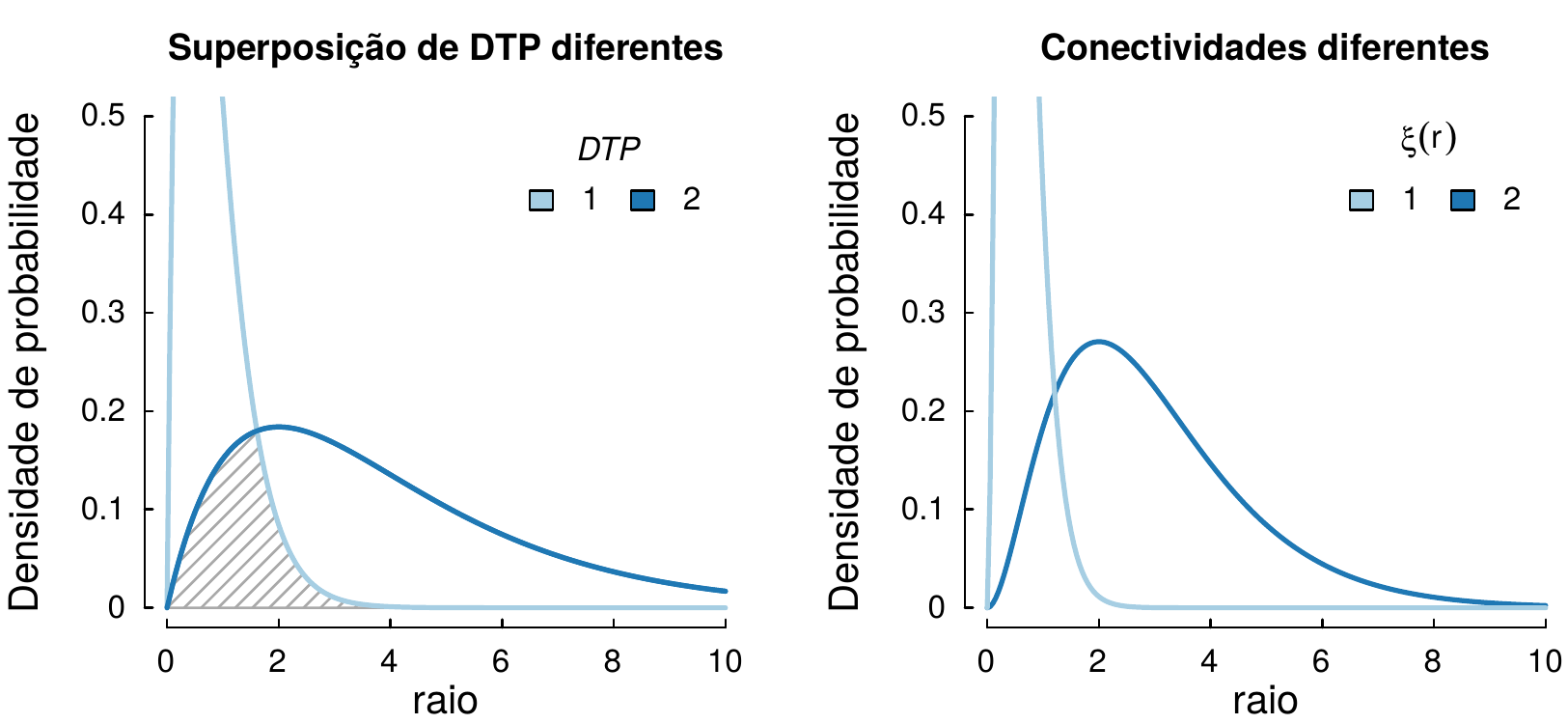}
        \caption{Example of two spherical pore size distribution overlapping and their respectives connectivity functions.}
        \label{Graf:SP}
    \end{figure}%
}

\newcommand{\grafDTPCarb}{%
    \begin{figure}[!h]
        \centering
        \includegraphics[keepaspectratio=true,width=0.95\textwidth]{\pastaGraf/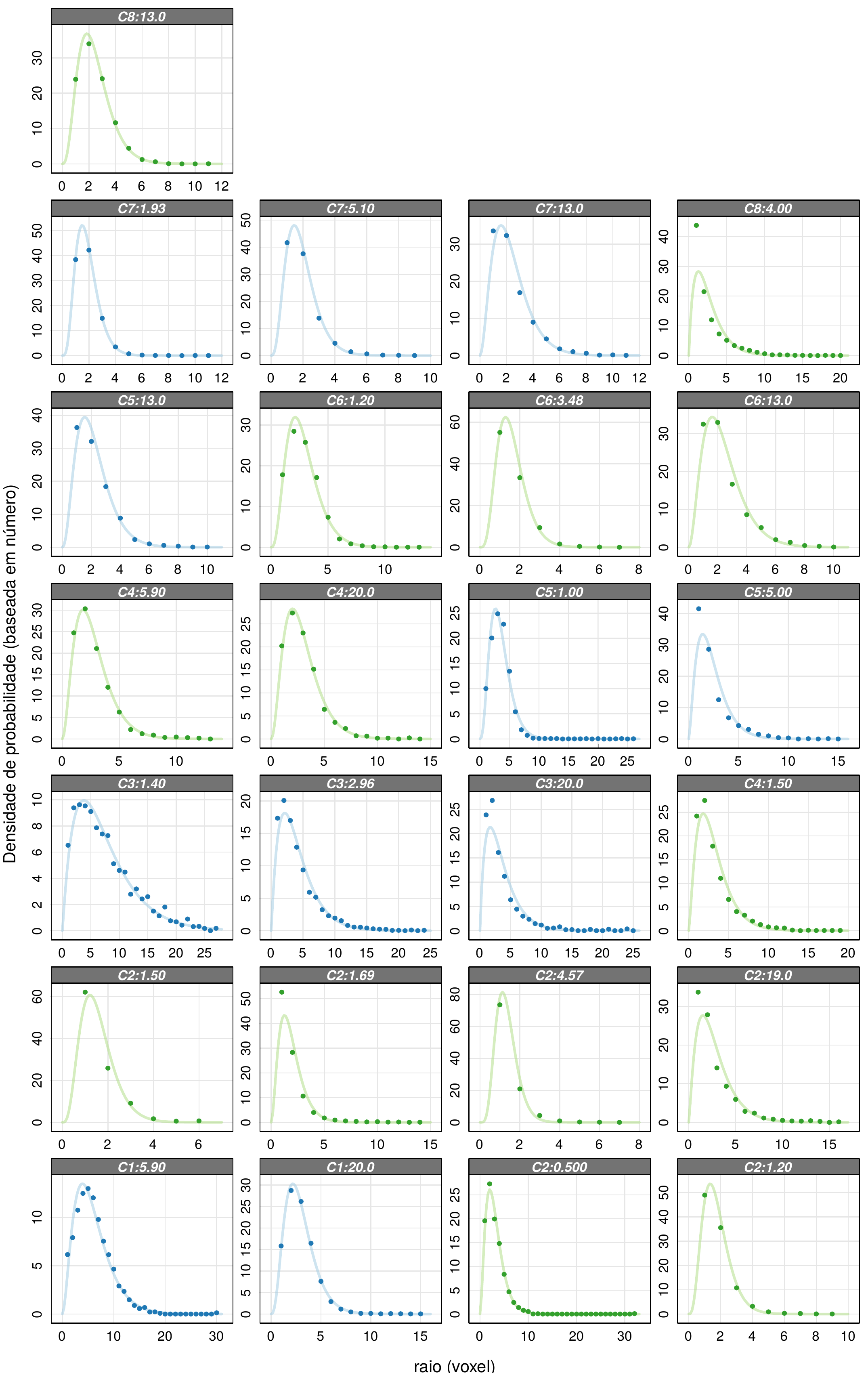}
        \caption{Spherical Pore Size Distribution (S-PSD) \Rr\ of carbonate samples fitted by the gamma distribution.}
        \label{Graf:DTPCarb}
    \end{figure}%
}

\newcommand{\grafCorrelCarb}{%
    \begin{figure}[!h]
        \centering
        \includegraphics[keepaspectratio=true,width=0.95\textwidth]{\pastaGraf/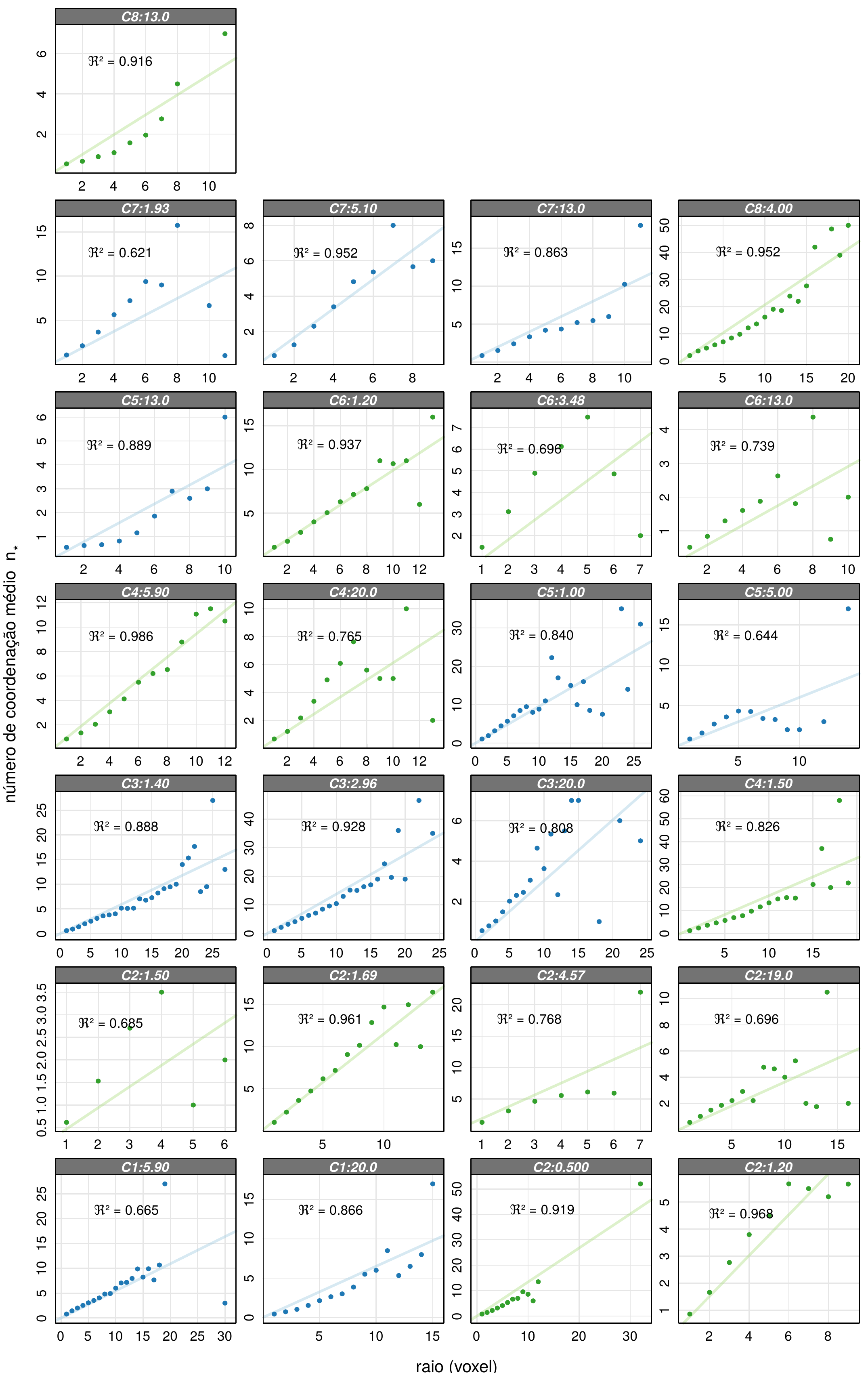}
        \caption{Correlation between mean coordination number $n\med$ of a pore and its radius $r$ for carbonate samples.}
        \label{Graf:CorrelCarb}
    \end{figure}%
}

\newcommand{\grafAjuste}{%
\begin{figure}[!h]
    \centering
    \subfloat[][]{{\includegraphics[keepaspectratio=true,width=0.85\textwidth]{\pastaGraf/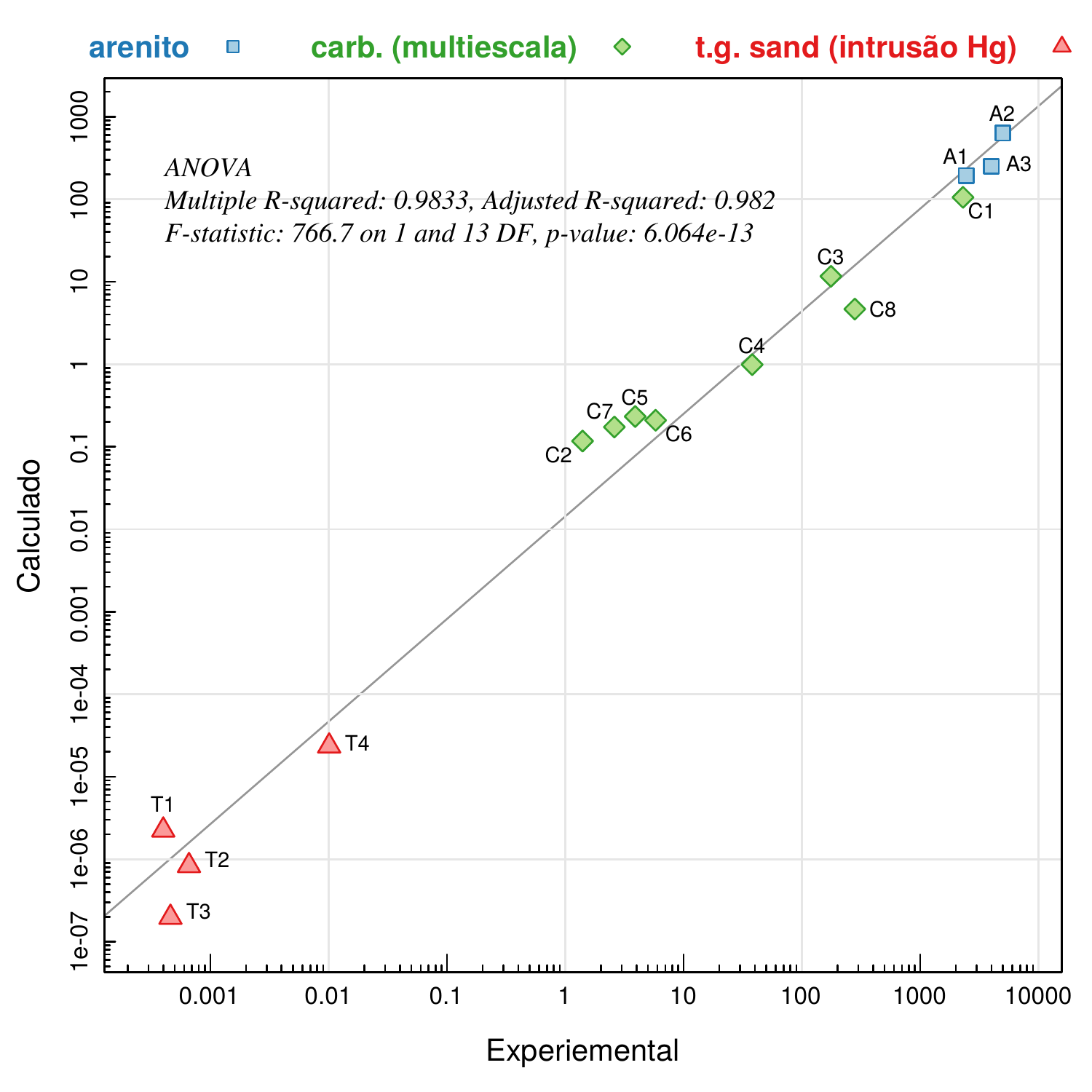}}
    \label{Graf:ComparaPerm}}%
    \\
    \subfloat[][]{{\includegraphics[keepaspectratio=true,height=5cm]{\pastaGraf/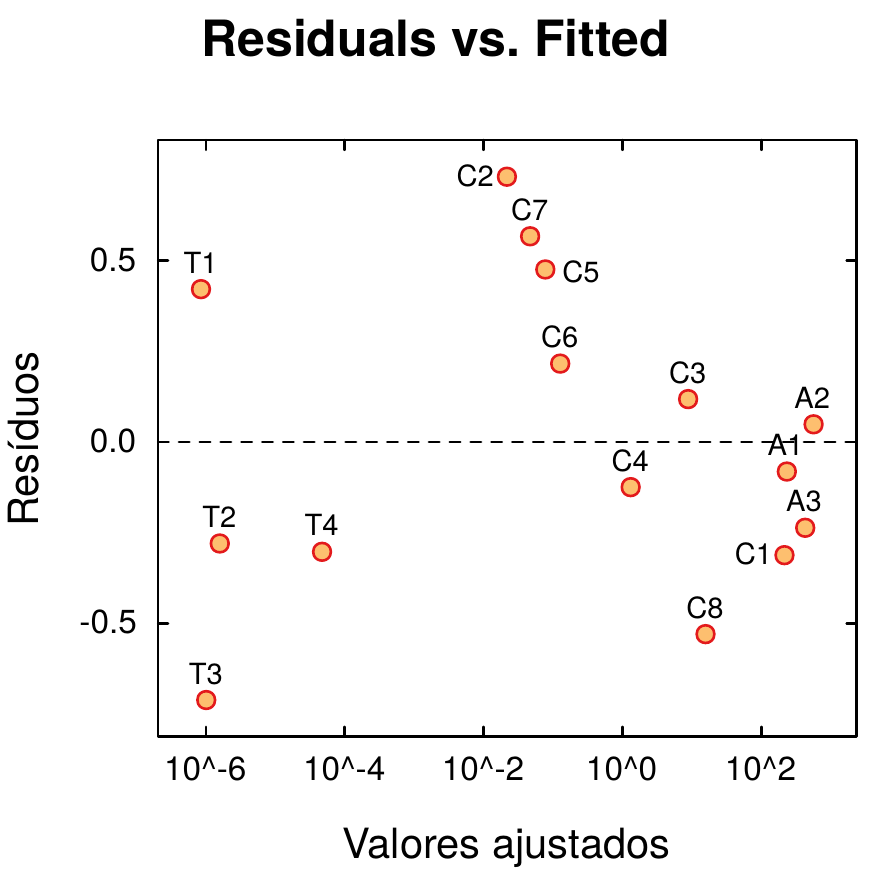}}
    \label{Graf:AjusteRes}}%
    \hfill%
    \subfloat[][]{{\includegraphics[keepaspectratio=true,height=5cm]{\pastaGraf/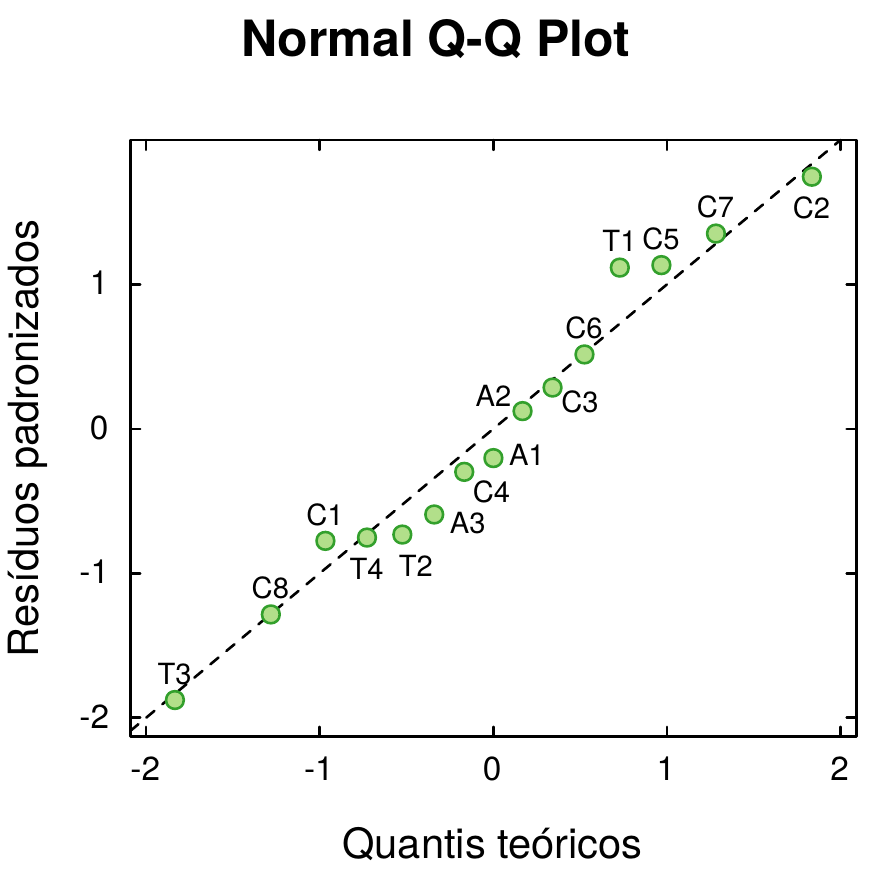}}
    \label{Graf:AjusteQQ}}
    \label{Graf:Ajuste}
    \caption{\sF{\protect\subref*{Graf:ComparaPerm}} Comparison between calculated and experimental permeability values. %
            \sF{\protect\subref*{Graf:AjusteRes}} Apparently random layout of the residuals.
            \sF{\protect\subref*{Graf:AjusteQQ}} Comparison of the standardized residuals with the theoretical quantiles.}
\end{figure}%
}

\newcolumntype{C}{>{\centering\arraybackslash}X}

\newcommand{\talpha}{$\alpha$}

\newcommand{\tp}{\normalfont\P}

\newcommand{\tabDadosGeral}{%
{%
\ctable[pos = !h,
        width = 0.95\textwidth,
        caption = Pore-throat network parameters., 
        label = Tab:DadosParam,
]{CcCCC}{
}%
{                                                                                                                                                                        \FL
    Sample                     & Resolution (\micro\metre) &     \talpha     &    $\theta$             &  \ensuremath{\Re^2}               \NN \cmidrule(lr){1-1} \cmidrule(lr){2-2} \cmidrule(lr){3-3} \cmidrule(lr){4-4} \cmidrule(lr){5-5}
    \plt A1                     &    \np{2.40}          &    \np{3.43}        &    \np{0.223}          &  \np{0.984}                       \NN
    \plt A2                     &    \np{3.40}          &    \np{3.05}        &    \np{0.119}          &  \np{0.978}                       \NN
    \plt A3                     &    \np{3.90}          &    \np{3.36}        &    \np{0.171}          &  \np{0.926}                       \NN
    \plt C1                     &    \np{5.90}          &    \np{2.90}        &    \np{8.25}\E{+04}    &  \np{0.665}                       \NN 
    \cellcolor{white} \mbox{}   &    \np{20.0}          &    \np{3.89}        &    \np{6.65}\E{+04}    &  \np{0.866}                       \NN
    \plt C2                     &   \np{0.500}          &    \np{2.92}        &    \np{1.90}\E{+06}    &  \np{0.919}                       \NN 
    \cellcolor{white} \mbox{}   &    \np{1.20}          &    \np{4.42}        &    \np{2.12}\E{+06}    &  \np{0.968}                       \NN 
    \cellcolor{white} \mbox{}   &    \np{1.50}          &    \np{4.56}        &    \np{1.96}\E{+06}    &  \np{0.685}                       \NN 
    \cellcolor{white} \mbox{}   &    \np{1.69}          &    \np{3.00}        &    \np{9.47}\E{+05}    &  \np{0.961}                       \NN 
    \cellcolor{white} \mbox{}   &    \np{4.57}          &    \np{6.46}        &    \np{1.06}\E{+06}    &  \np{0.768}                       \NN 
    \cellcolor{white} \mbox{}   &    \np{19.0}          &    \np{2.24}        &    \np{4.34}\E{+04}    &  \np{0.696}                       \NN
    \plt C3                     &    \np{1.40}          &    \np{2.04}        &    \np{1.96}\E{+05}    &  \np{0.888}                       \NN 
    \cellcolor{white} \mbox{}   &    \np{2.96}          &    \np{2.07}        &    \np{1.72}\E{+05}    &  \np{0.928}                       \NN 
    \cellcolor{white} \mbox{}   &    \np{20.0}          &    \np{1.94}        &    \np{2.83}\E{+04}    &  \np{0.808}                       \NN
    \plt C4                     &    \np{1.50}          &    \np{2.40}        &    \np{5.18}\E{+05}    &  \np{0.826}                       \NN 
    \cellcolor{white} \mbox{}   &    \np{5.90}          &    \np{3.03}        &    \np{1.90}\E{+05}    &  \np{0.986}                       \NN 
    \cellcolor{white} \mbox{}   &    \np{20.0}          &    \np{3.19}        &    \np{5.45}\E{+04}    &  \np{0.765}                       \NN
    \plt C5                     &    \np{1.00}          &    \np{4.07}        &    \np{1.17}\E{+06}    &  \np{0.840}                       \NN 
    \cellcolor{white} \mbox{}   &    \np{5.00}          &    \np{2.52}        &    \np{2.18}\E{+05}    &  \np{0.644}                       \NN 
    \cellcolor{white} \mbox{}   &    \np{13.0}          &    \np{3.48}        &    \np{1.24}\E{+05}    &  \np{0.889}                       \NN
    \plt C6                     &    \np{1.20}          &    \np{4.02}        &    \np{1.19}\E{+06}    &  \np{0.937}                       \NN 
    \cellcolor{white} \mbox{}   &    \np{3.48}          &    \np{5.20}        &    \np{9.40}\E{+05}    &  \np{0.696}                       \NN 
    \cellcolor{white} \mbox{}   &    \np{13.0}          &    \np{3.10}        &    \np{1.00}\E{+05}    &  \np{0.739}                       \NN
    \plt C7                     &    \np{1.93}          &    \np{4.97}        &    \np{1.38}\E{+06}    &  \np{0.621}                       \NN 
    \cellcolor{white} \mbox{}   &    \np{5.10}          &    \np{4.19}        &    \np{4.33}\E{+05}    &  \np{0.952}                       \NN 
    \cellcolor{white} \mbox{}   &    \np{13.0}          &    \np{3.09}        &    \np{1.02}\E{+05}    &  \np{0.863}                       \NN
    \plt C8                     &    \np{4.00}          &    \np{1.95}        &    \np{1.88}\E{+05}    &  \np{0.952}                       \NN 
    \cellcolor{white} \mbox{}   &    \np{13.0}          &    \np{4.06}        &    \np{1.28}\E{+05}    &  \np{0.916}                       \LL
}}}

\newcommand{\tabDadosGeralMerc}{%
{%
\ctable[pos = !h,
        width = 0.70\textwidth,
        captionskip = 0ex,
        cap = {Calculated parameters from the mercury intrusion curves},
        caption = Calculated parameters from the mercury intrusion curves., 
        label = Tab:DadosParamMerc,
]{CCC}{
}%
{                                                                        \FL
    Sample     &   \talpha         &    $\beta$             		     \NN \cmidrule(lr){1-1} \cmidrule(lr){2-2} \cmidrule(lr){3-3}
    \plt T1     &   \np{2.10}       & \np{5.92}\E{-08}                   \NN
    \plt T2     &   \np{0.527}      & \np{3.69}\E{-07}                   \NN
    \plt T3     &   \np{1.69}       & \np{3.43}\E{-08}                   \NN
    \plt T4     &   \np{2.14}       & \np{1.30}\E{-07}                   \LL
}}}

\newcommand{\tabPerm}{%
{\ctable[pos = !h,
        width = 0.80\textwidth,
        captionskip = 0ex,
        caption = Calculated permeability values.,
        label = Tab:Perm,
]{CCc}{%
\tnote[\tp]{Value obtained from the mercury intrusion curves.}%
}%
{                                                                                    \FL
 Sample    &   $k$ (mD)             &   Used resolution (\micro\metre)               \NN \cmidrule(lr){1-1} \cmidrule(lr){2-2} \cmidrule(lr){3-3} 
\plt  A1    &   \np{194}             &   \np{2.40}                                   \NN
\plt  A2    &   \np{634}             &   \np{3.40}                                   \NN
\plt  A3    &   \np{249}             &   \np{3.90}                                   \NN
\plt  C1    &   \np{1.70}\E{+03}     &   \np{5.90} ; \np{20.0}                       \NN
\plt  C2    &   \np{1.90}            &   \np{0.500}; \np{1.20}                       \NN 
\plt  C3    &    \np{188}            &   \np{1.40} ; \np{2.96}                       \NN
\plt  C4    &   \np{16.0}            &   \np{1.50} ; \np{5.90}                       \NN 
\plt  C5    &   \np{3.76}            &   \np{1.00} ; \np{13.0}                       \NN
\plt  C6    &   \np{3.39}            &   \np{1.20} ; \np{13.0}                       \NN 
\plt  C7    &   \np{2.80}            &   \np{1.93} ; \np{5.10} ; \np{13.0}           \NN
\plt  C8    &   \np{75.3}            &   \np{4.00} ; \np{13.0}                       \NN
\plt  T1    &  \np{2.25}\E{-06}      &   \textit{\footnotesize Hg}\tmark[\tp]          \NN
\plt  T2    &  \np{8.34}\E{-07}      &   \textit{\footnotesize Hg}          \NN   
\plt  T3    &  \np{1.97}\E{-07}      &   \textit{\footnotesize Hg}          \NN
\plt  T4    &  \np{2.35}\E{-05}      &   \textit{\footnotesize Hg}          \LL
}}}

\begin{document}

%
%

\title{A phenomenological connectivity measure for the pore space of rocks%
\footnote{Research supported by Brazilian agencies CAPES, CNPq and FAPESC, and Petrobras.}
}

\author{%
    Andr\'e Rafael Cunha\footnote{A. R. Cunha. Porous Media and Thermophysical Properties Laboratory, Federal University of Santa Catarina, SC, Brazil. \url{andre.cunha@posgrad.ufsc.br}.}
    \and
    Celso Peres Fernandes
    \and
    Lu\'is Orlando Emerich dos Santos
    \and
    Denise Prado Kronbauer
    \and
    Iara Frangiotti Mantovani
    \and
    Anderson Camargo Moreira
    \and
    Mayka Schmitt
}

\maketitle

\begin{abstract}
The interconnectivity of the porous space is an important characteristic in the study of porous media and their transport properties.
Hence we propose a way to quantify it and relate it with the intrinsic permeability of rocks.
We propose a measure of connectivity based on geometric and topological information of pore-throat network, which are models built from microtomographic images, and we obtain an analytical method to compute that property.
The method is expanded to handle rocks that present a higher degree of heterogenity in the porous space, which characterization requires images from different resolutions (multiscale analysis).
Trying to expand the methodology beyond the scope of images, we also propose a new interpretation for the experiment that generates the mercury intrusion curve and calculate the permeability.
The methodology was applied to images of 11 rocks, 3 sandstone and 8 carbonate rock samples, and to the experimental mercury intrusion curve of 4 tight gas sand rock samples.
We observe as result the existence of a correlation between the experimental and the predicted values.
The notions of connectivity developed in this work seek above all to characterize a porous material before a typical macroscopic phenomenology.

\medskip

Pore space connectivity.
Transport properties.
Intrinsic permeability.
Microtomographic images of rocks.
Multiscale analysis.
Mercury intrusion curve.

\medskip


\end{abstract}

%
%

\section{Introduction}
\label{intro}

The porous space does not have a regular geometry.
Nevertheless, it is usual to assign to it someone to allow a mathematical treatment.
The most common ones are the capilar tubes and the networks models.

The first modeling attempts admitted that the porous medium is formed by capillary tubes \citep{Kozeny1927,Carman1937}.
The application of physical laws to these models is facilitated by the simple geometry.
By definition, these models do not contemplate porous space connectivity.
Therefore its application is limited to certain classes of materials \citep{Scheidegger1963}.

An alternative idea is to consider the pore space as a network formed by \textit{pores}, larger spaces that store fluid, and \textit{throats}, which restrict the flow while performing the communication between the pores \citep{Dullien1979}.
Under this view, two quantities are relevant for the displacement of matter: the radius of the pore and the number of throats that leave (or reach) that pore.
The first is of geometric nature, and the second, topological.
The number of throats of a pore is called \textit{coordination number} of that pore.

The representation ways of the porous medium by network are related to the development of computation, 
since a network is formed by many constituents,
and due to imaging techniques, which can provide information from the material.
In the 1950s, some authors used to circumvented the problem of excessive calculations by means of electromechanical analogies 
(\citealp{Bruce1943}, cited by \citealp{Scheidegger1963}; \citealp{Owen1952}, cited by \citealp{Sahimi1993}; \citealp{Fatt1956a,Fatt1956b,Fatt1956c}).
At an intermediate stage,
The 2D imaging techniques allowed the introduction of images to the simulation.
But a 2D image is not able to adequately represent porous space connectivity (\citealp{ChatzisDullien1977}, cited by \citealp{VanMarcke2010}).
Therefore, criteria were developed to generate new random networks with statistical image information%
\footnote{They are named pixel/voxel based statistics or point-to-point statistics.}%
,
which are superimposed to build a 3D volume where the phenomenon is simulated. 
And even if higher-order statistics \citep{Okabe2005} or multiscalar schemes \citep{Fernandes1996} are considered, 
the generate volume does not properly express the real pore space.
The advent of X-rays microtomography technique in the porous media research in the 1980s \citep{Vinegar1987,Dunsmuir1991,BryantBlunt1992,Landisa2010}
made possible the observation of the real porous space complexity.
A phenomenon, such as a flow, can be simulated in a microtomographic image using some conventional numerical method \citep{VanMarcke2010}, 
But this procedure is computationally time consuming.
Therefore some simplification of the image is still tried,
and several methods return to the idea of a network \citep{AlKharusi2007},
but in this case the spatial configuration is represented with greater authenticity.
%


When geometrical shapes are signed to the pore space, it is called a \textit{pore-throat networks} or morphological networks,
where the phenomenum is described by the conservation laws, i.e., the \textit{continuum models}.
In \textit{random networks}, in turn, the phenomenon is approached by statistical physical theories, 
based on results of theories of percolation, renormalization, fractals and cellular automata, for example;
they are \textit{discrete models} \citep{Sahimi1993}.
In this work, we apply the Maximum Ball Algorithm \citep{HuDongPRE2009} to the microtomographic image.
The result is a network of spherical pores and cylindrical throats 
(Fig. \ref{Fig:ResultadoBM}).
In some case, the simulation based on the discretization of the motion equations is summarized to a linear system \citep{Cunha2013Physicae}.

\figResultadoBM
\subsection{Connectivity}

Although the idea of porous space connectivity is an intuitive truth, there is no single definition,
and the attempts to quantify it vary according to the branch of research.
In mathematics, connectivity is synonymous with \textit{topology} \citep{GrahamFlegg1974}.
Thus, the first attempt to describe it goes through topological definitions.
Therefore, in general, from a theoretical point of view, the coordination number has been the basis for quantifying connectivity
and generally the only parameter explored.
According to \citet{Sahimi1993}, Betti numbers are the most accurate way to characterize connectivity.
%
As the reference, we restrict ourselves to the first two numbers to ilustrate.
The first Betti number \betti{0} is the number of separate components that make up a structure.
A value $\betti{0}>1$ may indicate that the structure contains isolated porosity \citep{Sahimi1993}.
The second Betti number \betti{1} is the number of holes in a structure.
\betti{1} is equivalent to the \textit{genus} of a surface, which is the maximum number of closed curves that do not intercept and can be built on a surface without dividing it into distinct regions \citep{Vasconcelos1997}.
In this paper, we propose a phenomenological definition to connectivity as a conceptually less sophisticated alternative.
And before continue, we explore some examples from literature.

\citet{Vasconcelos1997,Vasconcelos1998}, for example,
adopting a 3D network model of cylindrical tubes, associates the genus per unit volume $G_V$ to the specific surface $S$ and the volume $V$, both experimentaly determined.
Then $G_V$ is introduced as a multiplicative factor directly into the permeability expression.

When dealing with random networks, the coordination number is an inherent and constant information of spatial organization.
Therefore, the more complex the network, the more it tends to be useful for describing real situations \citep{Efros1986}.
In morphological networks, in turn, the coordination number is not fixed,
and we can know a distribution.
\citet{Mason1982}, for example, interpreting the media as a random network, estimates the coordination number from adsorption isotherms.
For the author, connectivity is coordination number by definition,
and also comments on the limitation of considering a constant connectivity to what would probably be a distribution.

From the pixel based statistical view,
the connectivity of a random network can be defined from higher order moments of the phase function.
The intention is the 3D volume reconstruction.

Other examples of interpretation of connectivity in porous media are cited.
\citet{Glover2009} uses electrical parameters to propose a measure of connectivity, more precisely to the inverse of the resistivity of the rock formation.
\citet{Montaron2009}, in the same domain as the previous work, associates the connectivity of a random network to the conductivity equations obtained from percolation and medium field theories.
\citet{Trinchero2008},
in a groundwater perspective, consider the lack of a univocal concept for connectivity in this domain and adopt, as a measure of connectivity to an aquifer, the hydraulic response time between two points after the injection of markers in one.
\citet{Bernabe2010,Bernabe2011,Bernabe2016} define connectivity as the mean coordination number from pore-throat networks.

%
%
%

\subsubsection*{A new approach}

Considering a pore-throat network model, we propose a measure of connectivity in which the coordination number is not the only input data, the other is the radius of the pore.
In other words, topology is not the only relevant information for connectivity, so geometry is.
It is a phenomenological perspective, based on the weighting of the contribution of each object of the network to a flow.
What is more important: a large pore with low coordination number, or a small pore with high coordination number?
We propose that the interaction between the two informations can be the answer.
And from the quantification of the connectivity, we propose a quantitative correlation with the intrinsic permeability of the porous medium.

We start from the observation that the pore-throat networks exhibit interesting patterns:
the pore size distribution can be fitted by the gamma distribuition,
and the mean coordination number of a pore increases linearly with its radius.

\subsection{On the representative elementary volume (REV)}

It is very difficult to define an ideal volume size at which physical properties tend to stabilize,
and even if a certain physical property does, there are no guarantees that others will do so \citep{Dvorkin2009},
and as the larger scales are considered, it contributes with the increase in the physical-chemical heterogeneity of the porous formation.
In practice, the \textit{elementary volume} (which may or may not be representative) is determined by the limitations of the equipment used.
It is part of this work to assume that the patterns presented by the pore-throat network 
serve as criteria for determining an elementary volume that is representative before the phenomena in question, a monophasic flow.

%
%

%
\section{Materials and methods}

\subsection{Materials}

The images are:
3 sandstone rocks, named A1, A2 and A3,
observed with the respective resolutions of 2.40 \micro\metre, 3.40 \micro\metre\ e 3.90 \micro\metre,
from which a cubic volume of edges 300 voxels were cropped.
8 carbonate rocks, named sequentially from C1 to C8,
observed with the respective resolutions of 
5.90 and 20.0 \micro\metre\ to C1,
0.500, 1.20, 1.50, 1.69, 4.57 and 19.0 \micro\metre\ to C2,
1.40, 2.96 and 20.0 \micro\metre\ to C3,
1.50, 5.90 and 20.0 \micro\metre\ to C4,
1.00, 5.00 and 13.0 \micro\metre\ to C5,
1.20, 3.48 and 13.0 \micro\metre\ to C6,
1.93, 5.10 and 13.0 \micro\metre\ to C7,
4.00 and 13.0 \micro\metre\ to C8,
from which a cubic volume of edges 500 voxels were cropped.
The experimental permeability values are (in milliDarcy):
to A1, 2.45%
;
A2, 5.00;
A3, 4.00;
C1, 105;
C2, 0.117;
C3, 11.6;
C4, 0.987;
C5, 0.232;
C6, 0.209;
C7, 0.173;
C8, 4.65.

In a second moment,
we work with experimental data from mercury porosimetry.
They are 4 tight gas sand samples, named sequentially from T1 to T4.
The length $L$ of the samples follow respectivelly:
0.0340,
0.0325,
0.0315 and
0.0331 \metre;
and the diameter $D$:
0.0370,
0.0380,
0.0375 and
0.0380 \metre;
and the permeability ones:
4.00\e{-4},
6.60\e{-4},
4.60\e{-4}
e 1.01\e{-2} mD.

%
%
%

\subsection{Methods}

\subsubsection{The experimental data}

It is observed that the Spherical Pore Size Distribution (S-DTP) \Rr\ for the sandstones can be approximated by a gamma distribution
\begin{align}
    \Rr &= \frac{1}{\Gamma(\alpha)} \beta^{\alpha} r^{\alpha-1} e^{-\beta r} \quad, \quad r \geq 0 \quad.
    \label{Eq:rGama}
\end{align}
where $\alpha$ and $\beta$ the parameters of the distribution, and $\Gamma(x)$ is the gamma function.

%
\grafDTPAren
%

It is also observed the existence of a linear correlation between the mean coordianation number $n\med$ of a pore and its radius $r$ (Fig.~\ref{Graf:CorrelAren}), i.e.,
{\allowdisplaybreaks%
\begin{align*}
    n\med &\sim r \quad,    \\
    n\med &= a r + b \quad. 
\end{align*}}
Theoretically a pore of radius $r=0$ does not exist, and implies no connected throat, i.e., $b=0$.
Then,
\begin{align}
    n\med &= a r \quad.    \label{Eq:Correlacao}
\end{align}

%
\grafCorrelAren
%

The observed correlation implies that one can express $N\med(n\med)$ in terms of $R(r)$ \citep{Kay2005}:
{\allowdisplaybreaks%
\begin{align*}
    N\med(n\med) &= \frac{1}{a}\ R \left( \frac{n\med}{a} \right)  \quad.
\end{align*}}%

\subsubsection{A phenomenological connectivity}

Faced with a flow in the pore-throat network, two quantities are relevant for the mass displacement: the pore radius and its coordination number.
As said before, the first has a geometric nature and the second topological.
Highlighting the interaction between these entities of spatial configuration,
we define the connectivity function $\xi$ as
\begin{align}
    \xi(n,r) &\sim N(n)\ R(r) \quad, \label{Eq:Conectividade}
\end{align}
where $r$ is the pore radius with distribution $R$, and $n$ is the coordination number with distribuition $N$.
The expression can be rewritten for the mean coordination number $n\med$:
\begin{align}
    \xi(n\med,r) &\sim N\med(n\med)\ R(r) \quad. \label{Eq:ConectividadeMedia}
\end{align}
{\allowdisplaybreaks%
The observed linear correlation, $n\med \sim r$, allows to rewrite the connectivity $\xi$ as a function only of radius $r$,
\begin{align}
    \xi(r) &\sim R(r)^2 \quad. \label{Eq:XiSimRR}
\end{align}}%
That is, the connectivity function $\xi$, which covers geometric and topological informations of the network, is completely characterized by only one of the variables, $n\med$ or $r$.
In the case of eq.~({\ref{Eq:XiSimRR}}), the radius $r$ was chosen because it can be measured by different techniques.
Normalizing the equation%
\footnote{The normalization condition requires that: $\alpha>\sfrac{1}{2}$ e $\beta>0$.}%
,
\begin{align}
    \xi(r) &= \frac{2^{2 \alpha -1} \beta  \Gamma (\alpha )^2}{\Gamma (2 \alpha -1)}\ R(r)^2 \quad. \label{Eq:Normalizacao}
\end{align}

Fig.~\ref{Graf:Con} shows an example of $R(r)$ and its respective connectivity density $\xi(r)$ for $\alpha=2$ and $\beta=2$.
The \Rr\ curve shows that the smaller pores are more abundant than the larger ones.
The discussed correlation establishes, in turn, that the larger pores have more connections.
Finally, the $\xi(r)$ curve mediates these contributions and reveals the pores that most contribute to the network connectivity.

%
\grafCon
%

%
%
%

\subsubsection{Permeability $k$}

The permeability $k$ is given by \citep{Scheidegger1963}:
\begin{align}
    k &= - \frac{\eta L Q}{A \left( p_\text{out} - p_\text{in} \right)} \quad ,       \label{Eq:k}
\end{align}
where $\eta$ is the viscosity of the fluid, $L$ is the length of the material, $A$ is the area of the section, $p_\text{in}$ and $p_\text{out}$ are the pressures applied at the inlet and outlet ends, respectively, and $Q$ is the flow through $A$. 
The flow $Q$ is given by
\begin{align*}
    Q &= - \frac{1}{\Omega} \left( p_\text{out} - p_\text{in} \right) \quad ,
\end{align*}
where $\Omega$ is he hydraulic resistance of the porous medium.
Then, the eq. (\ref{Eq:k}) becomes
\begin{align}
    k &= \frac{L\eta}{A\Omega} \quad .       \label{Eq:k_novo}
\end{align}
%
%
In the right side, $L$, $A$ and $\Omega$ are all macroscopic ($\eta$ is a fluid property).
But the hydraulic resistance $\Omega$ is affected by the microscopic characteristics of the porous space.
We then consider $\Omega$ as a mean from microscopic informations.

Strictly speaking, the hydraulic resistance of a cell in the pore-throat network has two parts:
\begin{align*}
\Omega = \Omega_p + \Omega_g \quad ,
\end{align*}
where $\Omega_p$ is the contribution due to the spherical pores,
and $\Omega_g$ is due to cylindrical throats.

The connectivity $\xi(r)$ explicitly considers only the radius $r$ of the pores and the average number of throats that depart from that pore,
but not the geometry of those links.
However the Maximum Ball Algorithm establishes a relation between the geometries of the radius and its connected throat \citep{HuDongPRE2009, \bibCunhaMSA}. 
Then we can rewrite%
%
%
, 
\begin{align}
    \Omega  &= \Omega_p\ (1+\tau) \quad , \nonumber \\
    \Omega  &\propto \Omega_p \quad . \nonumber 
\end{align}%
And since the objective of this work is to demonstrate the existence of a correlation,
we write, without loss of generality,
\begin{align}
    \Omega  &= \Omega_p \quad , \nonumber
\end{align}
therefore \citep{\bibCunhaMSA}
\begin{align}
    \Omega  &= \frac{81 \eta}{\pi r^3} \quad . \label{Eq:Omega}
\end{align}

At this point the connectivity function $\xi$ is used to weight an expected value of $r^3$ in eq. (\ref{Eq:Omega}) \citep{Cunha2011Physicae}
{\allowdisplaybreaks%
\begin{align*}
    \langle r^3\rangle_\xi &= \int_{0}^\infty \xi(r)\ r^3\ dr \quad,    \\
    \langle r^3\rangle_\xi &= \frac{1}{4} \alpha  \left(4 \alpha ^2-1\right) \beta ^3  \quad.
\end{align*}}%

Replacing eq. (\ref{Eq:Omega}) in eq. (\ref{Eq:k_novo}),
\begin{align}
    k &= \frac{\pi L}{81 A} \; \frac{4}{\alpha  \left(4 \alpha ^2-1\right) \beta ^3}  \quad .       \label{Eq:k_final}
\end{align}

%
%

The described methodology has two main limitations.
The first is related to the anisotropy of the medium;
it is known that the permeability is a tensor entitity \citep{Scheidegger1954,Liakopoulos1965,Szabo1968,Durlofsky1991},
i.e., it depends on the flow direction;
in this work, however, we departs from a PSD characteristic of the volume,
and that's why cubic volumes are considered.
The second limitation is related to the normalization procedure of eq.~(\ref{Eq:Normalizacao}),
which can not be reached for $\alpha\le\sfrac{1}{2}$.

\subsubsection{Multiscalar Analysis}

For carbonates, which show a high degree of heterogeneity in their porous structure,
we propose to interpret the porous medium as a succession of the involved scales
in such a way that the equivalent hydraulic resistance is understood as a serial association of the resistances of each scale.
We means
\begin{align}
    \Omega_\mathrm{eq}  &= \sum_i \Omega_i = \frac{81 \eta}{\pi} \sum_i \frac{1}{\langle r_i^3\rangle} \quad , \label{Eq:OmegaMulti}
\end{align}
and
\begin{align*}
    L &= \sum_i L_i     \quad.
\end{align*}
where $i$ is the number of scales.

Observation at different scales results in distributions that overlap in some region (Fig. \ref{Graf:SP}).
This means that some pores have been counted more than once, and their contributions to the flow are overestimated.
Therefore, in the calculation of the permeability, we will avoid to consider very close resolutions
and consider the connectivity function $\xi$ to give the proper weight of the radius measured by each scale.

%
%
\grafSP
%


%
%
%
\subsubsection{Mercury Intrusion Curves}\label{Texto:DTPExp}\label{Citacao:Schmitt2013D}

A pore size distribution resulted from a mercury porosimetry is constructed from capillary tubes model through the Young-Laplace equation, \dtpYL,
\begin{align}
    p &= \frac{2\sigma \cos\theta}{\lambda} \quad . \label{Eq:YL}
\end{align}
This model does not consider connectivity by definition.
Therefore we propose to calculate \Rr\ from \dtpYL\ before estimates the permeability.

A porous medium is considered to be completely saturated by a fluid.
And for each pressure $p$ applied, we measure the cumulative volume $V$ that leaves the structure.
In the $n$-th measurement, the medium reached the irreducible saturation, i.e., the curve has $n$ points.
And through Young-Laplace equation, we calculate the $n$ values of radius $\Lambda_n = \{\lambda_1, \dots, \lambda_n\}$.

Let $(p_1,V_1)$ a point of the experimental curve.
The Young-Laplace equation associates $p_1$ to $\lambda_1$, meaning that all capillaries with radii greater than or equal to $ \lambda_1$ are accessible at this pressure.
Thus, the pressure $p_1$ can expel an amount of fluid from all connected spherical pores whose radii are greater than or equal to $\lambda_1$. 
Mathematically it is expressed:
\begin{align*}
    V_1 &= \frac{4\pi}{3} \int_{\lambda_1}^\infty \xi(r) r^3 dr \; 
        =\; \frac{4\pi}{3} \; \frac{\beta^3\; \Gamma\left(2\alpha+2, \frac{2\lambda}{\beta}\right)}{8\; \Gamma(2\alpha-1)} \quad.
\end{align*}

Analogously, high pressures can reach the capillaries with smaller radii.
Theoretically, when pressure $p_*\rightarrow\infty$, the radius $\lambda_*\rightarrow0$.
In this case all the capillaries are accessible to the pressure $p_*$, as long as they are connected.
Hence the total accumulated volume is
\begin{align*}
    V_* &= \frac{4\pi}{3} \int_{0}^\infty \xi(r) \; r^3 dr \;
         = \; \frac{4\pi}{3} \; \frac{\alpha \beta^3 (4\alpha^2-1)}{4} \quad.
\end{align*}

The cumulative volume density $v_1$ is given by the division of $V_1$ by $V_*$,
\begin{align}
    v_1(r)  &= \frac{\Gamma\left(2\alpha+2, \frac{2\lambda_1}{\beta}\right)}{2\; \Gamma(2\alpha-1) \; \alpha (4\alpha^2-1)} \quad. \label{Eq:V1}
\end{align}
The previous equation is one of the $n$ equations,
and is a transcendental equation of $\alpha$ and $\beta$,
and its numerical solution is presented in the \apendice{\ref{Ap:ET}}.
With the values of the parameters, we can go back to eq. (\ref{Eq:k_final}).

%
%

\section{Results and discussion}\label{Texto:ReD}

For the sandstone samples, the hypotheses have already been observed in Figs. \ref{Graf:DTPAren} and \ref{Graf:CorrelAren}.

For the carbonates ones, follow Figs. \ref{Graf:DTPCarb} and \ref{Graf:CorrelCarb}.
It is noted a weakening of the linear correlation for certain resolutions,
which are deprecated for the calculation of the permeability,
unless for C1, since they are the only ones available.
Here there is an implicit consideration:
the hypotheses that an elementary volume is representative when it presents the explored patterns, i.e., 
a gamma distribuition to S-PSD and the linear correlation between the pore radius and its mean coordination number.

Tab. \ref{Tab:DadosParam} and Tab \ref{Tab:DadosParamMerc} show the parameters obtained from the pore-throat networks
and mercury intrusion curves, respectively.

\grafDTPCarb
\clearpage

\grafCorrelCarb
\clearpage

\tabDadosGeral
\clearpage

\tabDadosGeralMerc

The calculated permeability values are given in Tab. \ref{Tab:Perm}.
The graphics of Fig. \ref{Graf:ComparaPerm} compare it with the experimental data,
showing the ANOVA regression analysis (5\%), which is justified by the graphics of Figs. \ref{Graf:AjusteRes} and \ref{Graf:AjusteQQ}.

It is observed, therefore, a clear correlation between the experimental permeability values and those estimated by the defined connectivity function.



\tabPerm

\grafAjuste

%
%
%

\section{Summary and conclusions}\label{Texto:CF}

In this paper we discuss some ideia related to the connectivity of porous media,
which is an important intrinsic feature in the study of the transport properties of those materials.
The main objetive was to quatify it and related it to the intrinsic permeability of rock samples.

We start from the observation that the pore-throat networks exhibit interesting patterns:
the pore size distribution can be fitted by the gamma distribuition,
and the mean coordination number of a pore increases linearly with its radius.

Our thesis was to suppose that both geometry and topology of the network are important for the mass displacement before a monophasic flow.
Then, we propose a phenomenological connectivity function 
\begin{align}
    \xi(n,r) &\sim N(n)\ R(r) \quad, \tag{\ref{Eq:Conectividade}}
\end{align}
that assumed the form
\begin{align}
    \xi(r) &= \frac{2^{2 \alpha -1} \beta  \Gamma (\alpha )^2}{\Gamma (2 \alpha -1)}\ R(r)^2 \quad. \tag{\ref{Eq:Normalizacao}}
\end{align}
That equation can quantify how a pore is connected only by its radius, which can be know by different experimental techniques.

We used it to calculate the permeability from a single network and for several networks coming from differente resolution images (multiscalar analysis).
We still extrapolate its use beyound the scope of images, proposing a new interpretation of the mercury intrusion curves.
Those expressions gave us a analytical formula for the intrinsic permeability,
which results are consistently correlated with the experimental values. 

Those results make us affirm that the defined connectivity function is a relevant entity before a monophasic flow.

During the study we have established two important characteristics to the pore space.
The first is that, after observe the above explored patterns in some microtomographic images,
we operate reciprocally and imposed them as quantitative criteria to evaluate if a volume is representative from its original material.
The second is the new interpretation to the experiment that generates the mercury intrusion curves and how to build the ideal PSD from the \dtpYL.

\subsection*{Perspectives}\label{Texto:Pers}

In a first moment, it is expected that the observed patterns can be explored in other rocks, even in other porous materials, and in other scales of observations.

Secondly, we imagine that the connectivity function $\xi$ can be applied to the other phenomena.
If no, we still expect that one can start from a generic 
\begin{align}
    \xi(n,r) &\sim f\left(\ N(n), R(r)\ \right) \quad
\end{align}
to propose alternative expressions most suitable.

%
%
\subsection{Acknowledgements}

We thank Prof. Carlos Appoloni for the valuable comments.

%
%

\appendix
\setcounter{section}{0}

\section{Resolution of the transcendental equation}\label{Ap:ET}

We search the values of $\alpha$ and $\beta$ that satisfy the eq.~(\ref{Eq:V1}).
Therefore an additional equation is required.
It is related to the expected value of gamma distribuition \citep{Kay2005}:
\begin{align}
    \alpha \beta &= \langle r\rangle_R = \int_0^\infty R(r)\; r\; dr \quad .\label{Eq:Media}
\end{align}
None of the three terms in the preceding equation is still known.
Therefore, the experimental data DTP-YL is used to approximate the right side of eq.~(\ref{Eq:Media}).
To emphasize the central tendency of values, we choose to use the median of the set $\Lambda_n$, denoted by $\Lambda_M$.
We write
\begin{align}
    \alpha \beta &= \Lambda_M \quad. \label{Eq:Mediana}
\end{align}
An implicit consideration of the previous equation is that the DTP-YL experimental curve must also be close to a gamma distribution.
We can now replace 
\begin{align}
    \beta &= \frac{\Lambda_M}{\alpha} \quad, \label{Eq:AlfaBeta}
\end{align}
in the eq.~(\ref{Eq:V1}), and have a transcendental equation only for $\alpha$, whose solution can be searched numerically.

Ideally the parameters should be unique for the existing $n$ equations;
But in practice only $m < n$ equations have a solution.
And the value of the parameter will be the mean of the set of $m$ elements.

Since $\alpha$ is known, we return to eq. (\ref{Eq:AlfaBeta}) to determine $\beta$.
And now we are able to know \Rr\ and $\xi(r)$.

%
%

\bibliographystyle{plainnat}       
\bibliography{bib}   

\end{document}